\documentclass{aastex631}
\usepackage{savesym}
\savesymbol{tablenum}
\usepackage{siunitx}
\restoresymbol{SIX}{tablenum}
\usepackage{amsmath}
\usepackage{xcolor}
\usepackage{multirow}
\usepackage{threeparttable}

\newcommand{\ie}{\textit{i}.\textit{e}.}
\newcommand{\eg}{\textit{e}.\textit{g}.}

\newcommand{\Eqn}[1]{equation~(\ref{#1})}
\newcommand{\Fig}[1]{Figure~\ref{#1}}
\newcommand{\Tab}[1]{Table~\ref{#1}}
\def\um{\mbox{$\mu$m}}

\begin{document}
\title{Modeling the Wavelength Dependence of Pixel Response Non-Uniformity of A CCD Sensor}

\author{Zun Luo}
\affiliation{Key Laboratory of Space Astronomy and Technology, National Astronomical Observatories, Chinese Academy of Sciences, Beijing, 100101, China}
\affiliation{University of Chinese Academy of Sciences, Beijing, 101408, China}

\author[0000-0001-9781-6863]{Wei Du}
\altaffiliation{duwei@shnu.edu.cn}
\affiliation{Shanghai Key Laboratory for Astrophysics, Shanghai Normal University, Shanghai, 200234, China}
\affiliation{Key Laboratory of Space Astronomy and Technology, National Astronomical Observatories, Chinese Academy of Sciences, Beijing, 100101, China}

\author{Baocun Chen}
\affiliation{Key Laboratory of Space Astronomy and Technology, National Astronomical Observatories, Chinese Academy of Sciences, Beijing, 100101, China}

\author{Xianmin Meng}
\affiliation{Key Laboratory of Space Astronomy and Technology, National Astronomical Observatories, Chinese Academy of Sciences, Beijing, 100101, China}

\author[0000-0003-1718-6481]{Hu Zhan}
\altaffiliation{zhanhu@nao.cas.cn}
\affiliation{Key Laboratory of Space Astronomy and Technology, National Astronomical Observatories, Chinese Academy of Sciences, Beijing, 100101, China}
\affiliation{Kavli Institute for Astronomy and Astrophysics, Peking University, Beijing, 100871, China}

\begin{abstract}
Precision measurements in astronomy require stringent control of systematics such as those arising from imperfect correction of sensor effects. 
In this work, we develop a parametric method to model the wavelength dependence of pixel response non-uniformity (PRNU) for a laser-annealed backside-illuminated charge-coupled device. The model  accurately reproduces the PRNU patterns of flat-field images taken at nine wavelengths from 290nm to 950nm, leaving the root mean square (RMS) residuals no more than 0.2\% in most cases. 
By removing the large-scale non-uniformity in the flat fields, the RMS residuals are further reduced. 
This model fitting approach gives more accurate predictions of the PRNU than cubic-spline interpolation does with fewer free parameters.
It can be applied to make PRNU corrections for individual objects according to their spectral energy distribution to reduce the photometry errors caused by the wavelength-dependent PRNU, if sub-percent level precision is required.

\end{abstract}
\section{Introduction} \label{sec:section1}

Pixel response non-uniformity (PRNU) characterizes spatial variations of an image sensor's response to uniform illumination and can introduce non-negligible uncertainties to precision measurements in astronomy \citep[\eg,][]{4436363,Stubbs_2014,Tyson_2015}. Projects like the Dark Energy Survey  \citep[DES;][]{thedarkenergysurveycollaboration2005dark} and the Legacy Survey of Space and Time \citep[LSST;][]{LSST_overview_2019} have made a great effort to improve our understanding of PRNU patterns and other sensor effects as well as their impact on photometry, astrometry, shape measurements and cosmological parameters \citep{2014PASP..126..750P,2015JInst..10C5027B,2016ApJ...825...61O,Bernstein_2017,2017PASP..129h4502B,2018SPIE10709E..1LB,2020ApJ...889..182P}.
The PRNU of a CCD can be determined by taking flat-field images. 
In such images, one may identify patterns like brick walls \citep{1998SPIE.3355..598W,2014SPIE.9154E..16V,2015Ap&SS.358...47M} and tree rings \citep{1997eiad.conf.....M,2014PASP..126..750P,2017JInst..12C5015P}, which can stretch over hundreds of pixels or more.  

In general, shorter-wavelength photons generate electrons closer to the back surface of  backside-illuminated (BSI) CCDs, so PRNU patterns and their effects may exhibit various degrees of dependency on wavelength  \citep{2014PASP..126..750P,2014SPIE.9154E..16V,2015JInst..10C6004M}.
For instance, the brick-wall patterns are more pronounced in shorter wavebands for laser-annealed BSI CCDs \citep[e.g.,][]{2014SPIE.9154E..16V}, and the tree rings show similar behavior \citep[e.g.,][]{2017JInst..12C5015P}. 
The tree rings are caused by radial impurity variations arising during the growth of silicon ingot, which then induce electric fields affecting the collection of electrons \citep{Holland_2014,2014PASP..126..750P,2015JInst..10C5013A,2017JInst..12C4018B,Magnier_2018}.
Therefore, we expect the tree-ring pattern to vary gradually with wavelength as it is a cumulative effect of lateral perturbations along the photoelectron's path toward the front surface. The brick walls, on the other hand, are an imprint of the laser annealing process that stamps over and passivates the back surface of the CCD after ion implantation to improve quantum efficiency (QE) in the ultraviolet \citep{1998SPIE.3355..598W}.
Since ultraviolet photons are absorbed within tens of nanometers in silicon, the brick-wall patterns must be generated at the very top layer of the back surface and fade rapidly above $\sim 400\,\mathrm{nm}$ as the photon absorption length quickly exceeds the depth affected by the ion implantation and laser annealing processes.

One can predict PRNU patterns of a specific CCD by simulating the 3D transport of photoelectrons. It requires detailed knowledge of the distribution of impurities and defects in the CCD besides the CCD design information to solve for the electric field  \citep{10.1063/5.0058894}. However, a direct measurement of such properties in each pixel is challenging for large-format CCDs and has to be done with a destructive process.
With real observations, it is a common practice to model the PRNU directly with dome flats and star flats. 
For example, a Principal Components Analysis (PCA) has been successfully applied to DES flat images, and the tree-ring pattern is readily visible in the first principal component \citep{2017PASP..129k4502B}. 
While it is worth exploring the PCA method in  wavelength space, we keep our focus on a semi-physical model to describe the wavelength-dependent PRNU.

Experimenting with a laser-annealed BSI CCD, \citet{chen2018} found that its flat field could be fit well by a four-parameter semi-physical model (demonstrated with a small area of 300 pixels). The model assumes that the trapping probability of photoelectrons in the CCD decreases exponentially with the distance to the backside surface. Applying the model, \citet{2021arXiv210805674X} is able to reconstruct the flat fields at different wavelengths in a 50pix$\times$50pix cutout region of the JPAS-Pathfinder camera \citep{miniJPAS2021}, though their minimization of flat-field residuals
leads to parameter values distributed close to the initial input values.
Such problem might be caused by incomplete constraints or lack of sufficient regularization in the minimization process. In this paper, we apply the model to the entire frame (1024pix$\times$1024pix) of the same CCD studied in \citet{chen2018} and provide a robust fitting algorithm that finds the best-fit parameters independent of the initial guess.

The rest of this paper is organized as follows. In Section~\ref{sec:section2}, we describe the details about the experiments that are designed to measure the wavelength dependence of the CCD's PRNU. Section~\ref{sec:section3} introduces the PRNU model. Fitting method and PRNU reconstruction residuals are presented in Section~\ref{sec:section4}.
A potential application of our model fitting approach to  correct the photometric error caused by the wavelength-dependent PRNU is discussed in Section~\ref{sec:error}. 
Further discussion and conclusion of the results are given in Section~\ref{sec:summary}.

\section{Flat field data} \label{sec:section2}
The flat-field images are taken with an e2v CCD201-20 in an Andor iXon Ultra 888 camera. CCD201-20 is a  laser-annealed BSI electron multiplying CCD with $1024\times 1024$ active pixels. Its pixel size is 13~$\mu$m, and its thickness is roughly 13~$\mu$m (private communication with P.~Jerram from e2v). It is coated for highest sensitivity in the visible.
We set the camera temperature to -\SI{70}{\degreeCelsius} and operate the camera with the electron multiplying option turned off. 
Uniform illumination is obtained using an integrating sphere with LEDs as the light source. 
The central wavelength and full width at half-maximum (FWHM) of each LED's spectrum  are listed in Table~\ref{tab:LEDs} along with the corresponding photon absorption length $L$ in Silicon at -\SI{70}{\degreeCelsius}. The LEDs' spectral FWHMs are less than 20~nm, sufficiently narrow for our experiment.
The absorption length is first estimated using the absorption coefficient model given by \citet{1979SSEle..22..793R} and then calibrated by the experimental results at 300K from \citet{Green1995}.

\setlength{\parskip}{12pt}
\begin{deluxetable}{ccccccccccc}[ht!]\label{tab:LEDs}
\tablewidth{0pt} 
\tablehead{
 \colhead{LED} & \colhead{LED1} & \colhead{LED2} & \colhead{LED3} & \colhead{LED4} & \colhead{LED5}  & \colhead{LED6}  & \colhead{LED7} & \colhead{LED8} & \colhead{LED9}} 
\tablecaption{LED central wavelengths $\lambda_\mathrm{c}$, spectral FWHMs and corresponding photon absorption length $L$ in Silicon at -\SI{70}{\degreeCelsius}.
}
\startdata
 $\lambda_\mathrm{c}$ (nm) & 287& 309 & 366 & 467 & 591 & 632 & 726 & 850 & 947\\
FWHM (nm) & 18.0& 19.3& 9.5& 16.3& 13.1& 12.6& 15.5& 15.3& 14.8 \\
 $L$ ($\mu$m) & 0.0048 & 0.0075 & 0.0123 & 0.8161 & 3.1929 & 4.4547 & 9.3180 & 29.897 & 113.04\\
\enddata
\end{deluxetable}

For each pixel in the flat fields, its signal-to-noise ratio (SNR) is given by
\begin{equation}
\mathrm{SNR}=\frac{N_\mathrm{p}}{\sqrt{N_\mathrm{p}+N_\mathrm{d}+RN^2}},
\end{equation}
where $N_\mathrm{p}$ and $N_\mathrm{d}$  are the numbers of photoelectrons and dark current electrons in the pixel, respectively, and $RN=4.7~\mathrm{e^-}$ is the readout noise. 
The CCD has a very low dark current, producing less than $0.1~\mathrm{e^-}$ within the maximum integration time of 74s. 
We adjust the integration time for each LED to collect around 35~ke$^-$ per pixel ($\sim 44\%$ full well), so that the SNR is dominated by the photon noise. One hundred exposures are then taken with each LED and are stacked to suppress the statistical noise in each pixel to less than $0.07\%$ ($\mathrm{SNR} > 1500$). 

The CCD camera used in our tests is equipped with an iris shutter. The shutter opens from the center outward and closes in reverse direction. This causes the exposure time of the pixels in the middle of the image to be longer than those at the edge. To account for this effect, we take a frame of 0.1s exposure (referred to as shutter-effect exposure) after each flat-field exposure. The final stacked flat field is then
\begin{equation}
I = \frac{1}{100}\sum_{k=1}^{100}(I_{F,k}-I_{S,k}),
\end{equation}
where $I_{F,k}$ and $I_{S,k}$ are pixel values of the $k$-th flat-field and shutter-effect exposures, respectively, and the pixel index and the wavelength label are omitted for simplicity. 

\Fig{fig:fig1} presents the normalized flat field (NFF) for the nine LED wavebands as defined by $f=I/\langle I\rangle$, where $\langle \ldots \rangle$ denotes an average over all pixels.
A striking feature is the brick-wall patterns that are prominent in the near ultraviolet (NUV) and fade toward longer wavelengths. 
The tree-ring patterns are also discernible at lower levels. As seen in Table~\ref{tab:LEDs}, NUV photons are absorbed within tens of nanometers from the back surface. Therefore, to form strong patterns in the NUV while keeping the visible and NIR only slightly affected, one needs a mechanism to remove or trap photoelectrons very close to the back surface of the CCD.
It is known this kind of trapping is responsible for low QEs in the ultraviolet\citep{1985SPIE..570....7J,Leach_1987,1991SPIE.1447..156H}. 
Processes such as ion implantation followed by laser annealing are necessary to boost the QE in the ultraviolet, and these processes can leave an imprint in the PRNU as we seen in Figure \ref{fig:fig1}. 

\section{Modeling} \label{sec:section3}
The PRNU model in \citet{chen2018} is derived from an effective model of QE and incorporates the absorption of photon and trapping of photoelectrons at the back surface of the CCD. In this section, we introduce the QE model first and then give the flat-field expression. Since our aim is to reconstruct and remove the PRNU rather than calculate the signal in each pixel, the model here actually describes the NFF instead of the absolute flat field. 

\subsection{Quantum efficiency model}\label{subsec:section3.1}
QE is defined as the ratio of detected electrons to incident photons. If we neglect the loss of electrons in the process of charge transfer and readout, the QE of a pixel can be written as \citep{1985SPIE..570....7J,Janesick..2007}
\begin{equation}
\eta = \alpha \mathcal{T}(1-e^{-H/L})\mathcal{C}\label{eq2}
\end{equation}
where $\alpha$ is the quantum yield (\ie\ the number of photoelectrons generated by an absorbed photon), $\mathcal{T}$ is the surface transmission, $H$ is the CCD thickness, $L$ is the photon absorption length, and $\mathcal{C}$ is the charge collection efficiency (CCE) (\ie\ the ratio between collected and generated photoelectrons). Note that $\alpha$, $\mathcal{T}$ and $L$, are wavelength dependent. The quantum yield $\alpha$ is greater than one for wavelengths shorter than about 350~nm. The product $\mathcal{T}(1-e^{-H/L})$ represents the fraction of photons absorbed in the silicon.

The CCE is unity if there is no loss of electrons in the collection process. 
However, recombination and trapping of the photoelectrons can happen before they are collected. 
To account for the fact that the brick-wall patterns are stronger in shorter wavebands, we assume the integral absorption probability of photoelectrons generated at a depth $x$ from the back surface ($x=0$) to decay exponentially with $x$, \ie{}
\begin{equation}
P(x) = Pe^{-x/d}\label{eq3}
\end{equation}
where $0\le P\le 1$ and $d$ is a characteristic scale on the order of 100nm. 
Note that $P(x)$ is \emph{not} a probability density but an integral of the probability density from $x$ all the way through the rest of the silicon. 
In the photoactive region of a pixel, the number of photoelectrons $n_e(x)$ generated per unit depth at $x$ equals to the number of photons $n_\gamma(x)$ absorbed there times the quantum yield $\alpha$, \ie{}
\begin{equation}
n_e(x) = \alpha n_\gamma(x) = \alpha \frac{N_{\gamma,0}}{L} e^{-x/L}\label{eq4}
\end{equation}
where $N_{\gamma,0} \equiv \int_0^\infty n_\gamma(x) \mathrm{d}x$ is the number of photons going through the back surface. With equations~(\ref{eq3}) and (\ref{eq4}), the CCE can then be estimated by
\begin{equation}
\begin{split}
\mathcal{C} = 1 - \frac{\int_0^H{n_e(x)P(x)\mathrm{d}x}}{\int_0^H{n_e(x)\mathrm{d}x}}\\
=1-\frac{Pd}{L+d}\frac{1-e^{-H(L+d)/(Ld)}}{1-e^{-H/L}}\\
\approx 1-\frac{Pd}{L+d}\frac{1}{1-e^{-H/L}}
\end{split}
\label{eq5}
\end{equation}
where the term $\exp(-H(L+d)/(Ld))$ is dropped in the last line as $H\gg d$. Substituting \Eqn{eq5} into \Eqn{eq2}, one gets a simple expression for the QE
\begin{equation}
\eta = \alpha \mathcal{T}(1-e^{-H/L}-\frac{Pd}{L+d}). \label{eq6}
\end{equation}

\begin{figure}[ht!]
\plotone{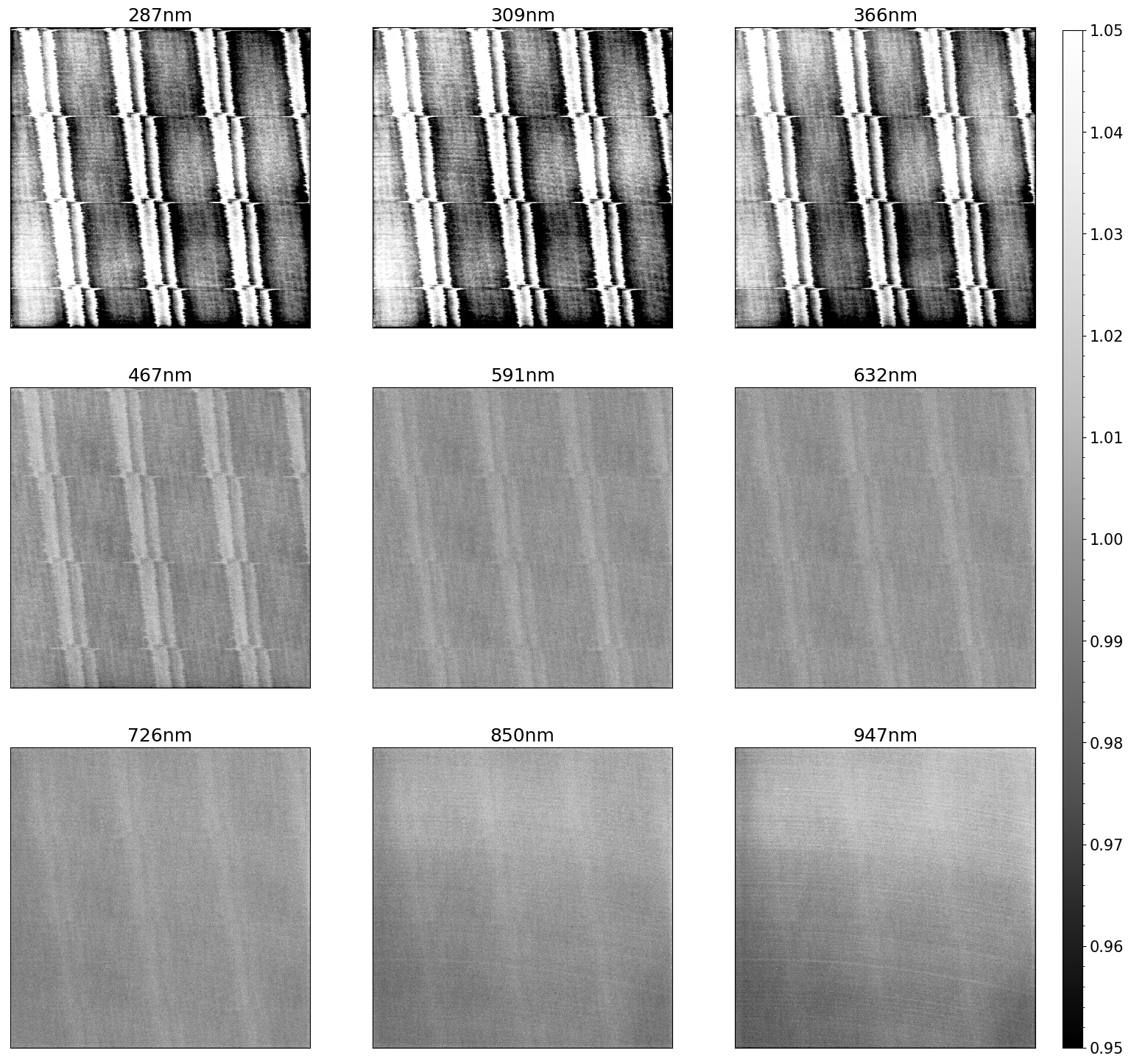}
\caption{NFFs ($1024\times 1024$ pixels) of nine wavebands as labelled above each panel. 
The periodic structures that are most prominent in the NUV are referred to as the brick-wall patterns.
The concentric rings that are barely discernible are known as the tree-ring patterns. 
The peak-to-valley value of the brick walls is about 18\% at 287nm and drops below 0.5\% at 947nm, while that of the tree rings decreases from about 1.6\% at 287nm to 0.7\% at 947nm.
To clearly show the PRNU structures at all wavelengths, we set the greyscale to be linear from black to white for pixel values from  0.95 to 1.05 in all the NFFs. 
\label{fig:fig1}}
\end{figure}

\begin{figure}[ht!]
\plotone{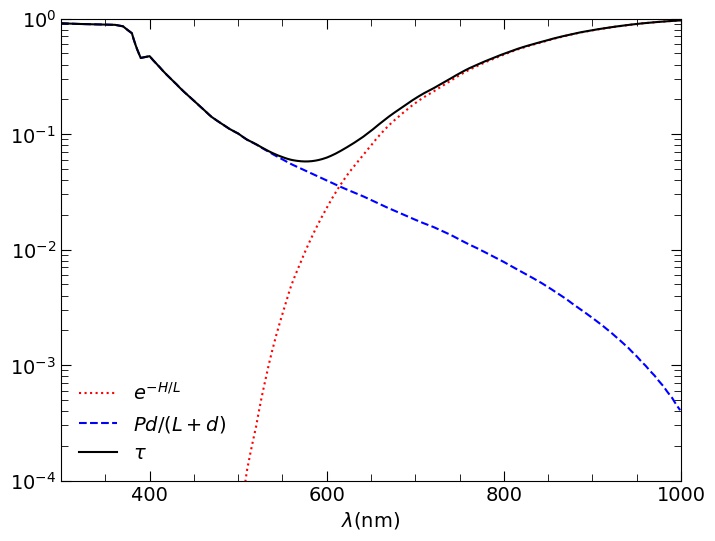}
\caption{Comparison of the photon absorption term $\exp(-H/L)$ (red dashed line) and the photoelectron absorption term  $Pd/(L+d)$ (blue dashed line). The sum of the two is shown in solid black line. The parameters $\left\{H, P, d\right\}$ assume the values of $\left\{13\um, 0.95, 0.15~\um{}\right\}$, and the photon absorption length in silicon $L$ is calculated at -\SI{70}{\degreeCelsius}. 
\label{fig:fig2}}
\end{figure}

\Fig{fig:fig2} compares the contributions of the terms $\exp(-H/L)$ and $Pd/(L+d)$ in \Eqn{eq6} with nominal values $H=13$~\um{}, $P=0.95$, and $d=0.15$~\um{}. 
The photoelectron absorption term $Pd/(L+d)$ is dominant at wavelengths $\lambda\lesssim 550$~nm as intended, whereas the CCD thickness becomes dominant at $\lambda\gtrsim 650$~nm. Hence, flat fields toward the blue side of the CCD response and those toward the red side are crucial for constraining the photoelectron absorption parameters $\{P,d\}$ and the CCD thickness $H$, respectively.

\begin{figure}[ht!]
\plotone{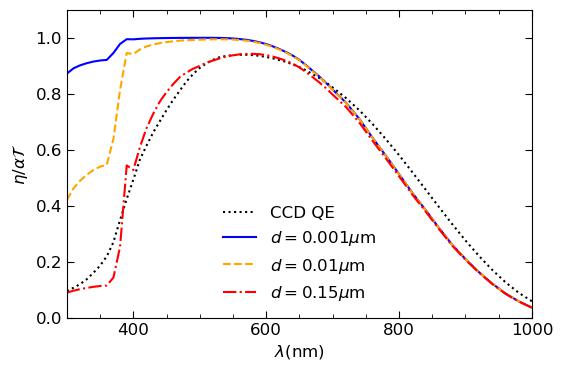}
\caption{Scaled QE (solid lines), \ie{} $\eta/(\alpha \mathcal{T})$, and the QE of the CCD at -\SI{70}{\degreeCelsius} (dotted line). The scaled QEs are calculated with the same $P = 0.95$ and $H = 13~\um{}$ but different $d$ values.
\label{fig:fig3}}
\end{figure}

\Fig{fig:fig3} illustrates the wavelength dependence of the QE scaled by $\alpha \mathcal{T}$. It is seen that the scaled QE is very sensitive to $d$ at short wavelengths ($P$ = 0.95 and $H = 13~\um$ held fixed). The larger the trapping scale $d$ is, the lower the QE will be.
Also shown in \Fig{fig:fig3} is the QE of the CCD (dotted line) that is not scaled by $\alpha \mathcal{T}$. 
Since the factor $\alpha \mathcal{T}$ is generally less than one in the wavelength range under discussion, the QE of the CCD are supposed to be lower than the scaled QEs at $\lambda\gtrsim650$~nm where $P$ and $d$ do not affect the QE. However, it is not the case, perhaps because we have not taken into account multiple reflections of low energy photons in the CCD. Multiple reflections can increase the chance of photons being absorbed and thus enhance the QE for these photons.
Neglecting such reflections could lead to an overestimation of $H$, which is not critical to our purpose.

\subsection{The PRNU model} \label{sec:FF_model}
The value $I_i$ in the $i$-th pixel can be written as
\begin{equation}
I_i = \frac{\eta_i A_i\mathcal{N}}{G} \label{eq7}
\end{equation}
where $G$ is the gain in $e^{-}$/ADU,  $\mathcal{N}$ is the number of incident photons per unit area, and $\eta_i$ and $A_i$ are the QE and area of the pixel, respectively. It has been noted that the pixel area can vary across the CCD because of irregular etching and doping \citep{4436363,10.1117/12.856519}. To proceed, we define
\begin{equation}
D_i(\lambda) = \alpha \mathcal{T}_i\frac{A_i\mathcal{N}}{G}. \label{eq8}
\end{equation}
Assuming the surface transmission $\mathcal{T}_i$ varies slowly across the CCD, we can decompose $D_i(\lambda)$ into a purely wavelength dependent component $D(\lambda)$ and a pixel dependent component, \ie{}
\begin{equation}
D_i(\lambda)\simeq D(\lambda)(1+\delta_i). \label{eq9}
\end{equation}
By definition, $\sum_i{\delta_i}=0$. Combining equations~(\ref{eq6}), (\ref{eq7}), (\ref{eq8}), and (\ref{eq9}), one gets
\begin{equation}
I_{i}(\lambda) = D(\lambda)(1+\delta_{i})\left(1-e^{-H_i/L}-\frac{P_{i}d_{i}}{L+d_{i}}\right).
\label{eq10}\end{equation}
Given that $D(\lambda)$ does not vary with pixels, the NFF $f_{i}$ can be reduced to
\begin{equation}
f_{i}(\lambda) = \frac{\tau_{i}}{\langle \tau_i \rangle} \label{eq11}\\
\end{equation}
with
\begin{equation}
\tau_{i}(\lambda) = (1+\delta_{i})\left[1-e^{-H(1+\Delta_i)/L}-\frac{P_{i}d_{i}}{L+d_{i}}\right], \label{eq12}
\end{equation}
where $H=13\um{}$ is redefined as the nominal thickness of the CCD, and $\Delta_i$ is the fractional difference of the thickness of the pixel with respect to $H$. It can be seen from equations~(\ref{eq11}) and (\ref{eq12}) that the NFF is a function of wavelength with the parameters $\{P_i,d_i,\Delta_i,\delta_i\}$. Moreover, the value of each pixel in the NFF depends not only on the four parameters of its own but also those of all other pixels ($1024\times 1024-1$ parameters in total).

\section{Flat field reconstruction}\label{sec:section4}
\subsection{Method}\label{subsec:section4.1}
We propose a two-step fitting procedure for better computational performance. First, we minimize the sum square residual of the NFF ratio (NFFR) $R_i(\lambda,\lambda_{\mathrm{ref}})$
\begin{equation}
R_{i}(\lambda,\lambda_{\mathrm{ref}}) \equiv \frac{I_i(\lambda)/I_i(\lambda_{\mathrm{ref}})}{\langle{I_i(\lambda)/I_i(\lambda_{\mathrm{ref}})}\rangle}=\frac{\tau_i(\lambda)/\tau_i(\lambda_{\mathrm{ref}})}{\langle{\tau_i(\lambda)/\tau_i(\lambda_{\mathrm{ref}})}\rangle} \label{eq13}
\end{equation}
\begin{equation}
\left\{ P_i,d_i,\Delta_i \right\} = \mathrm{argmin}[S_1(\left\{ P_i,d_i,\Delta_i \right\} )] \label{eq14}
\end{equation}
\begin{equation}
S_1\equiv \sum_k\sum_{j>k}{\sum_i{\left[R_{i}^{\mathrm{img}}(\lambda_j,\lambda_k)-R_{i}(\lambda_j,\lambda_k)\right]^2}}, \label{eq15}
\end{equation}
where the superscript ``$\mathrm{img}$'' denotes values calculated from real images and $\lambda_{\mathrm{ref}}$ is an arbitrary reference wavelength. The summation is over all pixels and all non-redundant NFFRs. 
Since the factor $(1+\delta_i)$ is cancelled in this step, the dimension of the optimization problem is reduced by 25\%.\\ 
\indent In the second step, a least square fitting of the NFFs is done to obtain $\left\{\delta_i \right\}$ with other parameters fixed to the results from the first step as follows
\begin{equation}
\left\{ \delta_i \right\} = \mathrm{argmin}[S_2(\{\delta_i\})]
\end{equation}
\begin{equation}
S_2\equiv \sum_j\sum_i\left[f_i^{\mathrm{img}}(\lambda_j)-f_i(\lambda_j)\right]^2_{\left\{ P_i,d_i,\Delta_i \right\}}.
\end{equation}

We make use of the L-BFGS-B constraint fitting routine in SciPy \citep{osti_204262}, which performs well on high dimensional nonlinear optimization problems with simple boundaries at low memory costs. It adopts a quasi-Newtonian line search method and needs an initial assignment of the parameters and the first-order derivatives of the target function, which can be found in Appendix \ref{sec:appendixA}. Fitting boundaries are set to $0.4<P<1$, $0.05\mu\mathrm{m}<d<0.2\mu\mathrm{m}$, $-0.3<\Delta<0.3$ and $-0.1<\delta<0.1$. The two-step minimization procedure for the fitting is validated with simulations in Appendix~\ref{sec:appendixB}. 

\begin{figure}[ht!]
\centering
\includegraphics[width=1.0\linewidth]{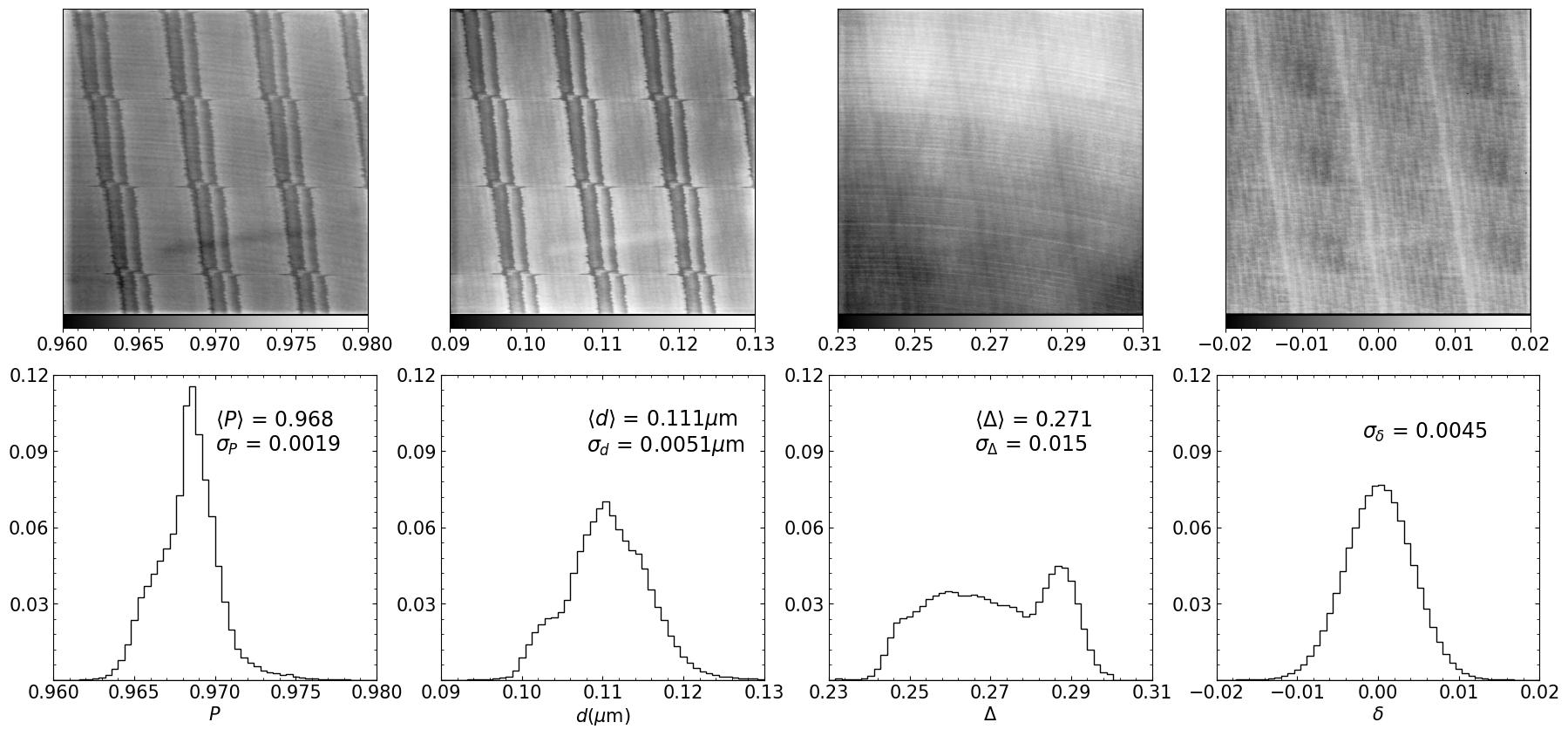}
\caption{Maps of the best-fit parameters $P$, $d$, $\Delta$, $\delta$ (top panels, $1024\times 1024$ pixels each) and their corresponding normalized histograms (bottom panels).
\label{fig:fig4}}
\end{figure}

\subsection{Direct reconstruction of the flat fields}\label{subsec:section4.2}
The flat-field images are fit with the procedure discussed in the last section. 
Given the huge dimension of the problem ($4\times10^6$) and its nonlinear nature, the result might trap in a local minimum or differ somewhat with a different initial guess due to accumulated numerical errors. 
We have carried out a test with randomly assigned starting parameter values in Appendix~\ref{sec:appendixB} and find that the ``best-fit'' parameters are nearly the same regardless of their starting point.

\Fig{fig:fig4} shows the maps of best-fit parameter values (top panels) and their corresponding normalized histograms. It is seen that the parameters $P$ and $d$ dominate the brick-wall patterns. Tree rings are identifiable in the maps of $P$ and $\Delta$. 
This is merely a result of model fitting, for $P$ and $\Delta$ are not related to the underlying cause of the tree rings. The histogram of $\delta$ is fairly Gaussian, whereas those for $P$, $d$ and $\Delta$ are clearly asymmetric. The RMS value of $\delta$ distribution is 0.45\%, consistent with 0.34\% and 0.88\% pixel size-area variations of two CCDs reported in  \citet{4436363}.
The relative deviation of the thickness $\Delta$ has an average of 0.271, consistent with the expectation from \Fig{fig:fig3} that $H$ could be overestimated. 

\begin{figure}[ht!]
\centering
\includegraphics[width=0.95\linewidth]{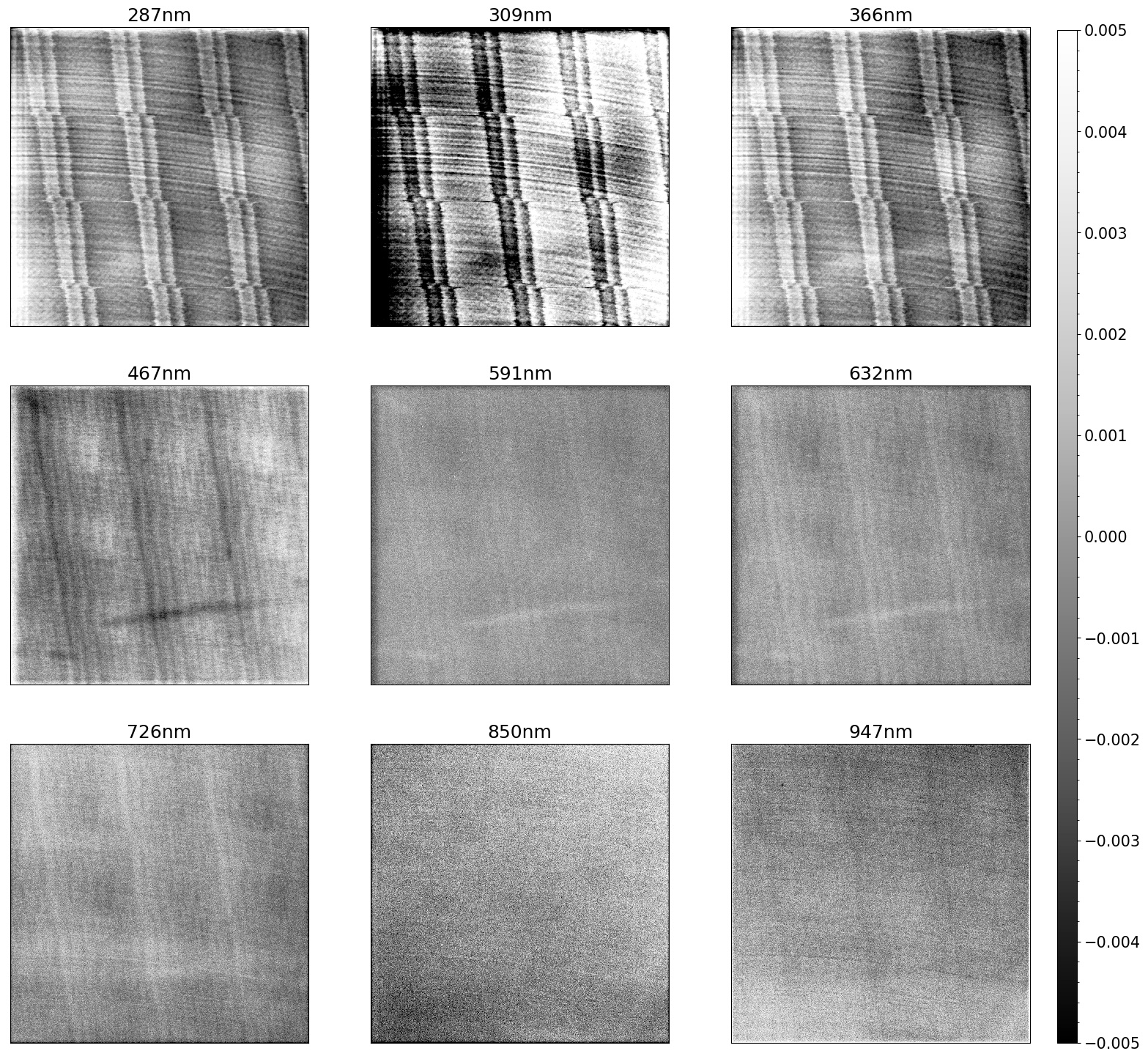}
\caption{Residuals of the best-fit model NFFs relative to
the observed NFFs in \Fig{fig:fig1}.
\label{fig:fig6}}
\end{figure}

With the PRNU model and best-fit parameters, one can generate realistic flat-field images at any wavelengths. In \Fig{fig:fig6}, we show the errors of the best-fit model NFFs relative to the observed NFFs in the 9 LED wavebands. The brick-wall patterns are still present in the errors but at much smaller amplitudes than those of the fluctuations in the observed NFFs in \Fig{fig:fig1}. The tree-ring patterns now stand out in the NUV residual maps and almost disappear in the NIR. This means that the tree rings in the NIR can be mimicked fairly well by thickness variations of the CCD, but for NUV photons, process near the back surface is not sufficient to produce the tree rings. 
In other words, the tree rings must be caused by an effect throughout the silicon. 
This corroborates the theory that the tree-ring patterns in the flat-field images are the result of lateral electric field in the CCD \citep{Holland_2014,2014PASP..126..750P}. Note that the same variations in the dopant that generate the lateral electric field may also produce broadened tree-ring patterns of charge diffusion variations \citep{Magnier_2018,2021MNRAS.501.1282J}.
Our model assumes the pixels to be independent of each other and, hence, does not account for the diffusion effect.

\begin{figure}[ht!]
\centering
\includegraphics[width=0.75\linewidth]{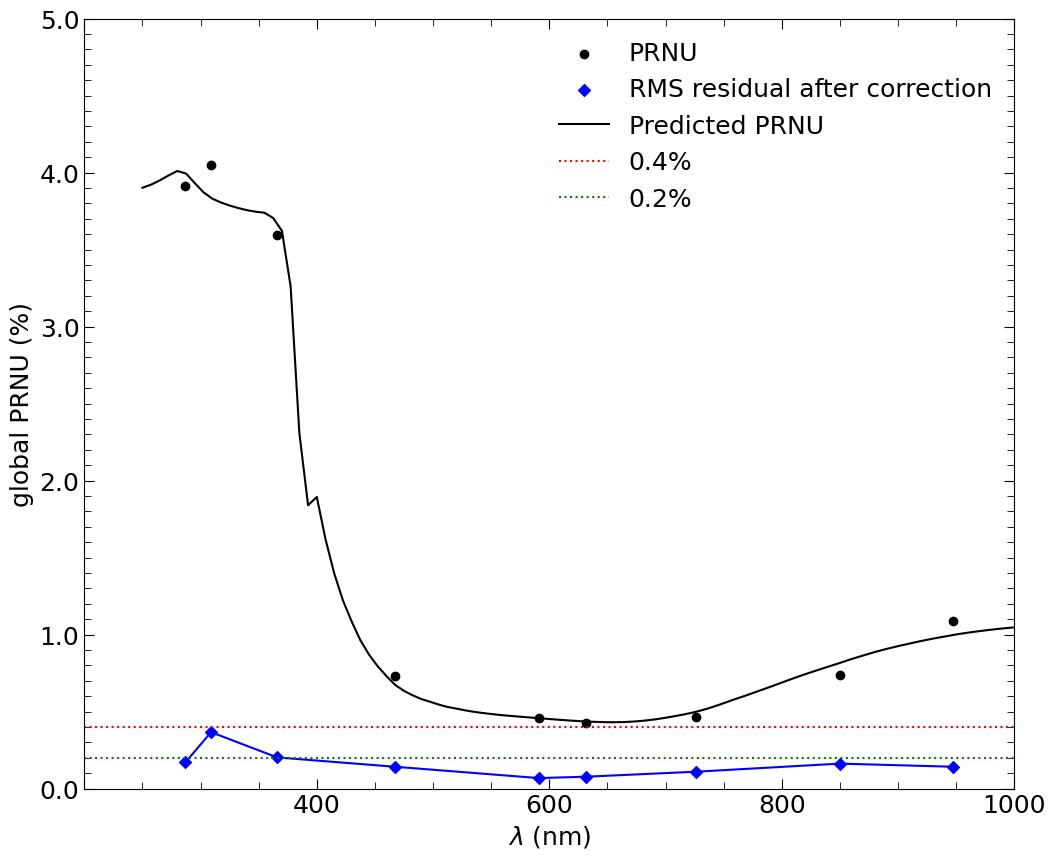}
\caption{The PRNU of the CCD (filled circles) and RMS residuals after correction using the best-fit model (filled diamonds). Prediction of the PRNU is drawn in a solid line. The horizontal dotted lines mark the value of 0.4\% and 0.2\%.
\label{fig:fig8}}
\end{figure}

\Fig{fig:fig8} presents the CCD's PRNU (solid circles) as quantified by the RMS of its NFFs. The black line is estimated from NFFs constructed with the best-fit parameter values. The filled diamonds correspond to the RMS values of the error maps in \Fig{fig:fig6}. The PRNU reaches 4\% in the NUV and drops quickly by an order of magnitude in the visible. 
The rise of the PRNU in the NIR to 1\% is mainly due to the large-scale feature of brightening in the upper half of the CCD as seen in \Fig{fig:fig1}. The RMS residuals are no more than 0.2\% except at 309nm, where it reaches 0.37\%. Around 600nm, the RMS residuals are close to the photon noise level of about 0.07\%. Based on the reconstruction test, we expect our method to be able to predict the PRNU accurately at any wavelength within the range sampled by the flat fields.

\begin{figure}[ht!]
\centering
\includegraphics[width=\linewidth]{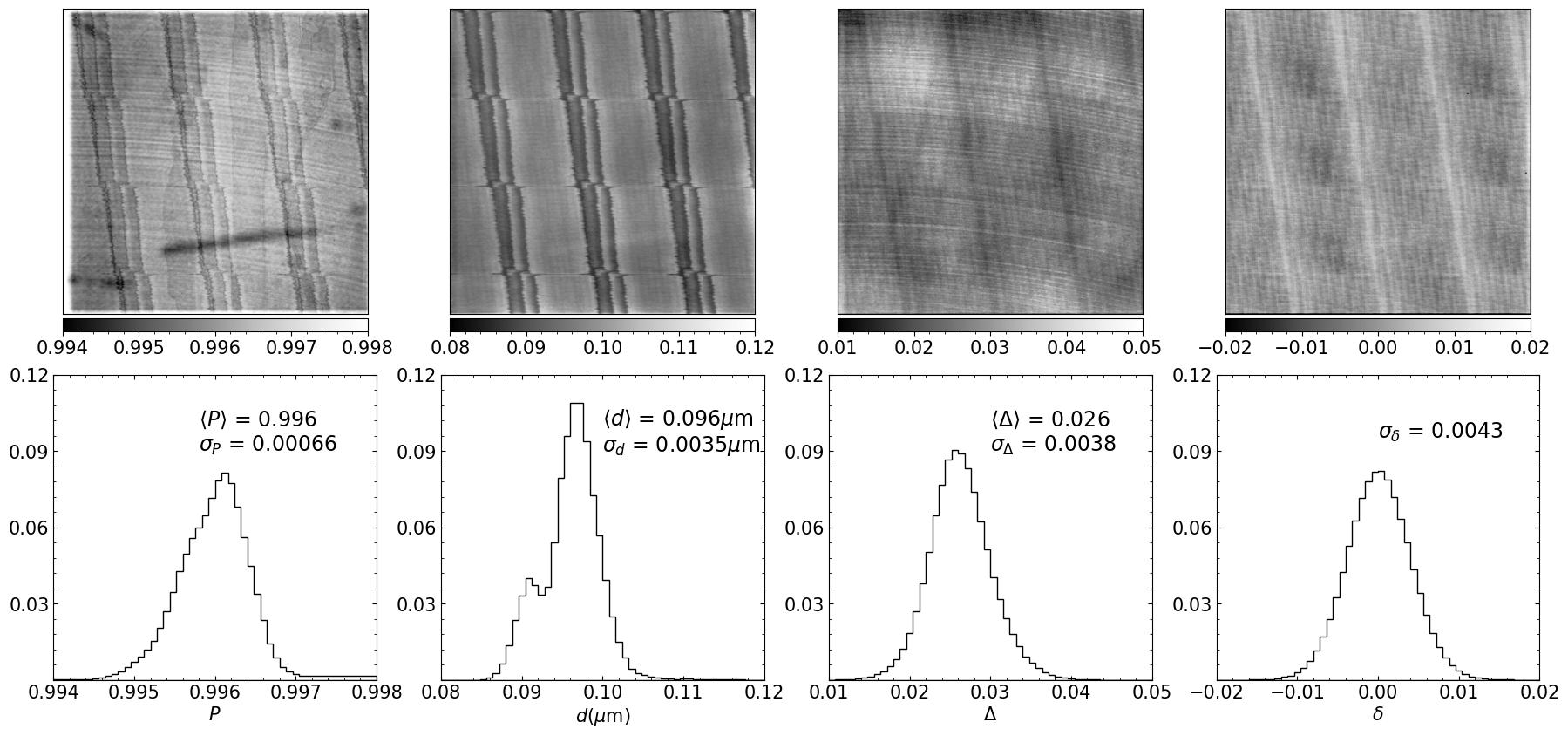}
\caption{Same as Figure \ref{fig:fig4} but with LSNU removed before parameter fitting.
\label{fig:fig11}}
\end{figure}

\subsection{Reconstruction with large-scale non-uniformity removed}\label{subsec:section4.3}

In the direct reconstruction, we have neglected non-uniformity of the light source. 
Since the distance between the integrating sphere and the CCD\ (about 2m) is far greater than the size of the CCD imaging area\ (19mm diagonal), we expect the illumination to vary only slowly across the CCD. 
To remove such large-scale non-uniformity (LSNU) from the PRNU in the previous subsection (hereafter, referred to as the total PRNU), we fit each flat field in Figure~\ref{fig:fig1} with a second-order two-dimensional polynomial and then divide each pixel value by that calculated from the polynomial. 
It should be noted that the intrinsic large-scale PRNU of the CCD is also removed in this way. 
Although the resulting images are no longer described by the physical model in Section~\ref{sec:section3}, we still apply the procedure in Section~\ref{subsec:section4.1} as an effective method to model and reconstruct the flat fields after removing the LSNU.

Figure \ref{fig:fig11} displays the best-fit parameter maps and histograms for the LSNU-removed NFFs. 
The features in the spatial distributions of the parameters are similar to those without removing the LSNU in Figure \ref{fig:fig4}. 
The scatters of the parameters become smaller, especially for $d$ and $\Delta$ that mainly affect the PRNU at NUV and NIR wavelengths, respectively. 
The most pronounced difference is that the departure of $H$ from its nominal value of 13\,\um{} is decreased from 27$\%$ to 2.6$\%$. This suggests that non-uniform illumination might have contributed to the inconsistency between the average value of the best-fit CCD thickness $H$ and its nominal value.

\begin{figure}[ht]
\centering
\includegraphics[width=0.75\linewidth]{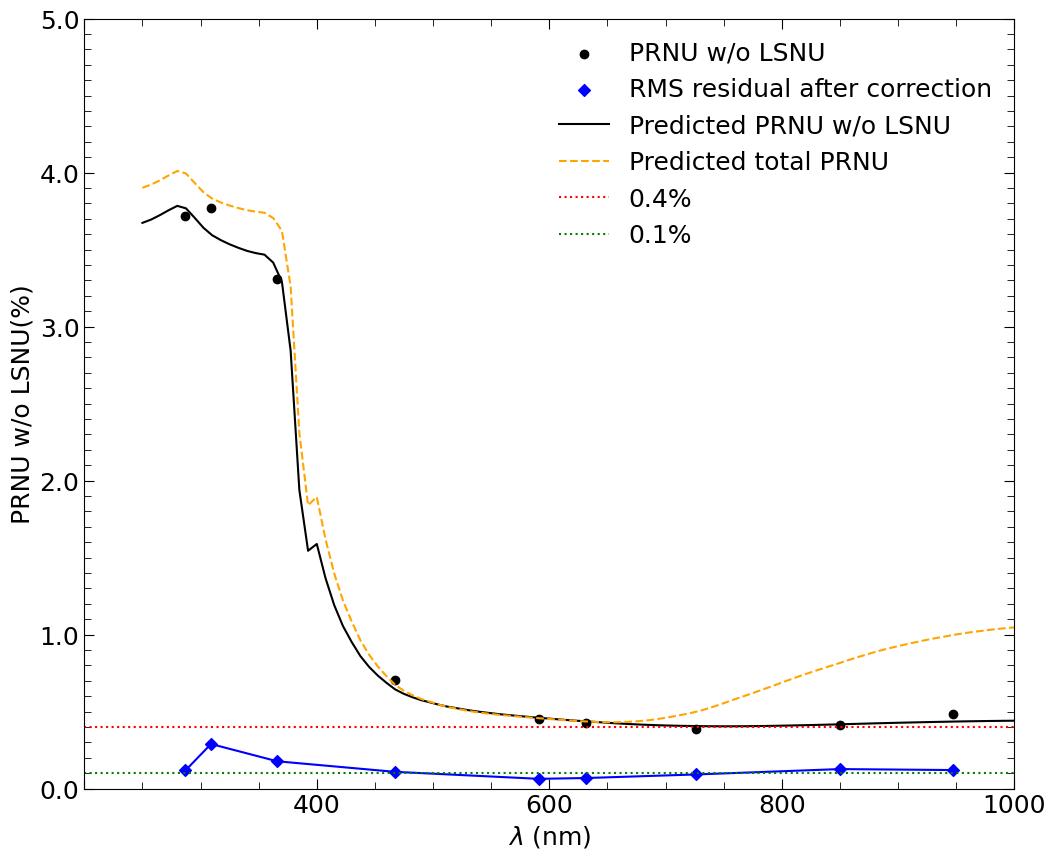}
\caption{Same as Figure~\ref{fig:fig8} but with LSNU removed before parameter fitting. The predicted total PRNU is added from Figure~\ref{fig:fig8} for comparison. 0.4\% and 0.1\% values are marked by horizontal dotted lines.
\label{fig:fig10}}
\end{figure}

The LSNU-removed PRNU is shown in Figure~\ref{fig:fig10}. 
One sees that the LSNU makes only a small contribution to the total PRNU at NUV and optical wavelengths, but contributes significantly to the total PRNU at NIR wavelengths. The RMS residuals are reduced from an average of 0.16\% before removing the LSNU to 0.13\% after removing the LSNU, and the RMS at most wavelengths is close to 0.1\%.

\subsection{Flat field prediction}\label{subsec:section4.4}
Results in Section \ref{subsec:section4.2} and Section \ref{subsec:section4.3} have demonstrated the ability of our model to accurately reconstruct the flat fields already taken. 
Here we check the performance of the model for interpolating flat fields at wavelengths that have not been imaged and compare the residuals with those by cubic-spline interpolation. Specifically, we remove one of the nine flat-field images at a time and apply our model fitting to the others. The model prediction of the NFF at the removed wavelength is then generated with the best-fit parameters and is subtracted from the real one to get the residuals.  

\begin{deluxetable}{cccccccccc}[ht!]
\tablewidth{0pt} 
\tablehead{ \colhead{LED} & \colhead{LED2} & \colhead{LED3} & \colhead{LED4} & \colhead{LED5} & \colhead{LED6} & \colhead{LED7} & \colhead{LED8}}
\tablecaption{RMS residuals of the PRNU model prediction (RMS$_\mathrm{m}$) and cubic spline interpolation (RMS$_\mathrm{s}$) of the flat field that is removed from the dataset.
\label{tab:interpolation}
}
\startdata
$\lambda_\mathrm{c}$(nm) & 309 & 366 & 467 & 591 & 632 & 726 & 850\\
RMS$_\mathrm{m}$(\%) & 0.51 & 0.83 & 0.23 & 0.09 & 0.10 & 0.12 & 0.22\\
RMS$_\mathrm{s}$(\%) & 0.50 & 0.83 & 1.32 & 0.24 & 0.13 & 0.17 & 0.35\\
\enddata
\end{deluxetable}
In \Tab{tab:interpolation}, we list the RMS values of the residuals for the PRNU model (RMS$_\mathrm{m}$) and cubic-spline interpolation (RMS$_\mathrm{s}$).  
It is worth noting that the model uses only 4 parameters to describe the flat-field behavior across the wavelength range, whereas the cubic-spline interpolation needs 28 coefficients to do the same. 
Nevertheless, the PRNU model clearly outperforms the other method. 
The difference between the two diminishes in the NUV because there is not sufficient NUV data to determine the NUV-sensitive parameters accurately. 
We expect the PRNU model to improve with additional flat fields imaged at more NUV wavelengths.

\section{Photometry error from wavelength-dependent PRNU} \label{sec:error}
The wavelength-dependent PRNU adds an error in photometry calibration if the spectral energy distributions (SEDs) of the target object, the standard star and the flat-field illumination source are not matched. 
Suppose that a target star with an SED $S^{\mathrm{t}}$ is centered at the $j$-th pixel of a CCD (the pixel index is collapsed to one dimension for convenience). The pixel-averaged sensor QE is $\langle \eta_i(\lambda) \rangle$, and the transmission in the band observed is $T(\lambda)$. The electron counts $N_{ij}^{\mathrm{t}}$ per second in pixel $i$ can be expressed as
\begin{equation}
\begin{split}
N_{ij}^{\mathrm{t}} = p_{ij}\int T(\lambda)S^{\mathrm{t}}(\lambda)\langle\eta_i(\lambda)\rangle\frac{\tau_i(\lambda)}{\langle\tau_i(\lambda)\rangle}\lambda\mathrm{d}\lambda = F^{\mathrm{t}}C^{\mathrm{t}}p_{ij}n_{i}^{\mathrm{t}}, \label{eq18}
\end{split}
\end{equation}
\begin{equation}
F^{\mathrm{t}} \equiv \int{TS^{\mathrm{t}} \lambda\mathrm{d}\lambda}, \label{eq19}
\end{equation}
\begin{equation}
C^{\mathrm{t}} \equiv (F^\mathrm{t})^{-1}\int{ TS^\mathrm{t}\langle\eta_i\rangle \lambda\mathrm{d}\lambda},
\label{eq20}
\end{equation}
\begin{equation}
n_i^{\mathrm{t}} \equiv 
\frac{1}{F^\mathrm{t}C^\mathrm{t}}
\int{TS^\mathrm{t}\langle\eta_i\rangle\frac{\tau_i}{\langle\tau_i\rangle}}\lambda\mathrm{d}\lambda,
\label{eq21}
\end{equation}
where $p_{ij}$ is the point spread function (PSF) value in the $i$-th pixel ($\sum_i p_{ij}=1$), 
$F^{\mathrm{t}}$ is proportional to the total in-band photon counting rate of the target star, 
$C^\mathrm{t}$ is the SED-weighted average QE in the band,
and $n^\mathrm{t}_i$ is equivalent to an NFF in the band using $S^\mathrm{t}$ as the light source's SED ($\langle n_i^\mathrm{t} \rangle=1$). 

A constant factor has been dropped in \Eqn{eq18} for convenience,
and contributions such as the sky background, bias and dark 
current are also neglected without losing generality. 
The total electron counts per second of the target star is then
\begin{equation}
    N_j^\mathrm{t}=\sum_i N_{ij}^\mathrm{t}
    =F^\mathrm{t}C^\mathrm{t}\sum_i p_{ij}n_i^\mathrm{t}. \label{eq:Njt}
\end{equation}
Assuming that the PSF depends only on the displacement between the pixels $i$ and $j$, \ie{}, invariant within the field of view, one can obtain $N_j^\mathrm{t}$ (up to a factor of $F^\mathrm{t}C^\mathrm{t}$) by convolving $n_i^\mathrm{t}$ with the PSF and taking the value in pixel $j$. 

Now we use an illumination source with an SED $S^\mathrm{f}$ for flat fielding. The flat-field electron counts per second in pixel $i$ can 
be written similarly as
\begin{equation}
N_{i}^{\mathrm{f}}=\int{TS^\mathrm{f}\langle\eta_i\rangle\frac{\tau_i}{\langle\tau_i\rangle}}\lambda\mathrm{d}\lambda
=F^{\mathrm{f}}C^{\mathrm{f}}n_{i}^{\mathrm{f}},\label{eq23}
\end{equation}
\begin{equation}
F^\mathrm{f} \equiv \int{TS^{\mathrm{f}}
\lambda \mathrm{d}\lambda},\label{eq24}
\end{equation}
\begin{equation}
C^\mathrm{f} \equiv (F^\mathrm{f})^{-1}\int{ TS^\mathrm{f}\langle\eta_i\rangle\lambda\mathrm{d}\lambda},\label{eq25}
\end{equation}
\begin{equation}
n_i^{\mathrm{f}} \equiv
\frac{1}{F^\mathrm{f}C^\mathrm{f}}
\int{TS^\mathrm{f}\langle\eta_i\rangle\frac{\tau_i}{\langle\tau_i\rangle}}\lambda\mathrm{d}\lambda
=\frac{N_i^\mathrm{f}}{\langle N_i^\mathrm{f}\rangle}.
\label{eq26}
\end{equation}
The flat field corrected electron counts $N^{\mathrm{c,t}}_{ij}$ per second of the target star is
\begin{equation}
N^{\mathrm{c,t}}_{ij} = \frac{N^{\mathrm{t}}_{ij}}{n_i^{\mathrm{f}}}=
\frac{F^{\mathrm{t}}C^{\mathrm{t}}p_{ij}n^{\mathrm{t}}_{i}}{n_i^{\mathrm{f}}}. \label{eq27}
\end{equation}
For a standard star centered at the $k$-th pixel, we can calculate $n_{i}^{\mathrm{s}}$, $F^{\mathrm{s}}$, $C^\mathrm{s}$ and $N_{ik}^{\mathrm{c,s}}$  (the superscript $\mathrm{s}$ stands for the standard star) 
by replacing the target star's SED $S^\mathrm{t}$ with the standard star's SED $S^\mathrm{s}$ in equations~(\ref{eq18})--(\ref{eq21}).
Hereafter, $n_i^\mathrm{t}$, $n_i^\mathrm{s}$ and $n_i^\mathrm{f}$ are referred to as NFFs like $f_i$ in previous sections, even though they may or may not be derived directly from real images.
Note that the NFF $f_i$ is ideally monochromatic, whereas $n_i^\mathrm{f}$ is not necessarily so.

The magnitude of the target star $m^\mathrm{t}$ is obtained based on that of the standard star $m^\mathrm{s}$ via 
\begin{equation}
m^{\mathrm{t}} = m^{\mathrm{s}}+2.5\mathrm{log}\left(\frac{\sum_{i}{N^{\mathrm{c,s}}_{ik}}}{\sum_{i}{N^{\mathrm{c,t}}_{ij}}}\right) = m^{\mathrm{s}}+2.5\mathrm{log}\left(\frac{F^{\mathrm{s}}}{F^{\mathrm{t}}}\right)+\Delta m_{\mathrm{sys}}+2.5\mathrm{log}\left(\frac{\sum_i p_{ik} n_i^\mathrm{s}/n_i^\mathrm{f}}{\sum_i p_{ij} n_i^\mathrm{t}/n_i^\mathrm{f}}\right). \label{eq28}
\end{equation}
In principle, magnitudes should be calculated using fluxes instead 
of electron counting rates, though in practice one uses what
are registered in the pixels. 
In \Eqn{eq28}, the systematic error term $\Delta m_{\mathrm{sys}}\equiv 2.5\mathrm{log}\left(C^\mathrm{s}/C^\mathrm{t}\right)$ 
depends only on the pixel-averaged QE and SEDs of the target star and the standard star. 
It is usually dominant over the last term, which is an error varying with the locations of the target star and the standard star on the sensor.
Since our study focuses on the effects associated with the
wavelength-dependent PRNU, we list the results of
$\Delta m_{\mathrm{sys}}$ in \Tab{tab:magnitude_error} without further discussion.

The pixel dependent error in \Eqn{eq28} vanishes only if the target star, the standard star and the flat fields all have the same SED, or the two stars match perfectly in terms of both the SED and the position on the sensor (implying that the two are imaged at different times). 
It can also describe the residual of PRNU correction in time-domain observations by replacing the standard star with the target star itself. 
In such a case, the residual would be at the mmag level if the star is imaged randomly on the sensor ($\sigma_\mathrm{rep}$ in \Tab{tab:magnitude_error}). Therefore, observations requiring very high precision in repeatability, \eg, exoplanet detection by the transit method \citep{2016ASSL..428...89C,2018haex.bookE.117D}, are ideally done in space with the stars imaged at exactly the same positions on the sensor every time.

The RMS photometry calibration error $\sigma_{\mathrm{cal}}$ can be derived from \Eqn{eq28} by randomly sampling the positions of the stars over the whole sensor, which gives
\begin{eqnarray}
\sigma_{\mathrm{cal}} &=& 1.086\sqrt{\sigma_\mathrm{t,f}^2 + \sigma_\mathrm{s,f}^2},\label{eq29} \\
\sigma_\mathrm{x,f}^2 &=& \frac{\left\langle\left(\sum_i p_{il} n_i^\mathrm{x}/n_i^\mathrm{f}\right)^2\right\rangle}{\left\langle\sum_i p_{il} n_i^\mathrm{x}/n_i^\mathrm{f}\right\rangle^2}-1,
\nonumber
\end{eqnarray}
where the script x is either t for the target star or s for the standard star, and the system is assumed to be stable with time.
By setting $\sigma_\mathrm{s,f}=\sigma_\mathrm{t,f}$ in \Eqn{eq29}, one can obtain the RMS repeatability error of the target star over the whole sensor  $\sigma_\mathrm{rep}=1.536\sigma_\mathrm{t,f}$.
By construction, the repeatability error equals the calibration error  when the standard star's SED matches perfectly with the target star's SED. 

\begin{figure}[ht!]
\centering
\includegraphics[width=1.0\linewidth]{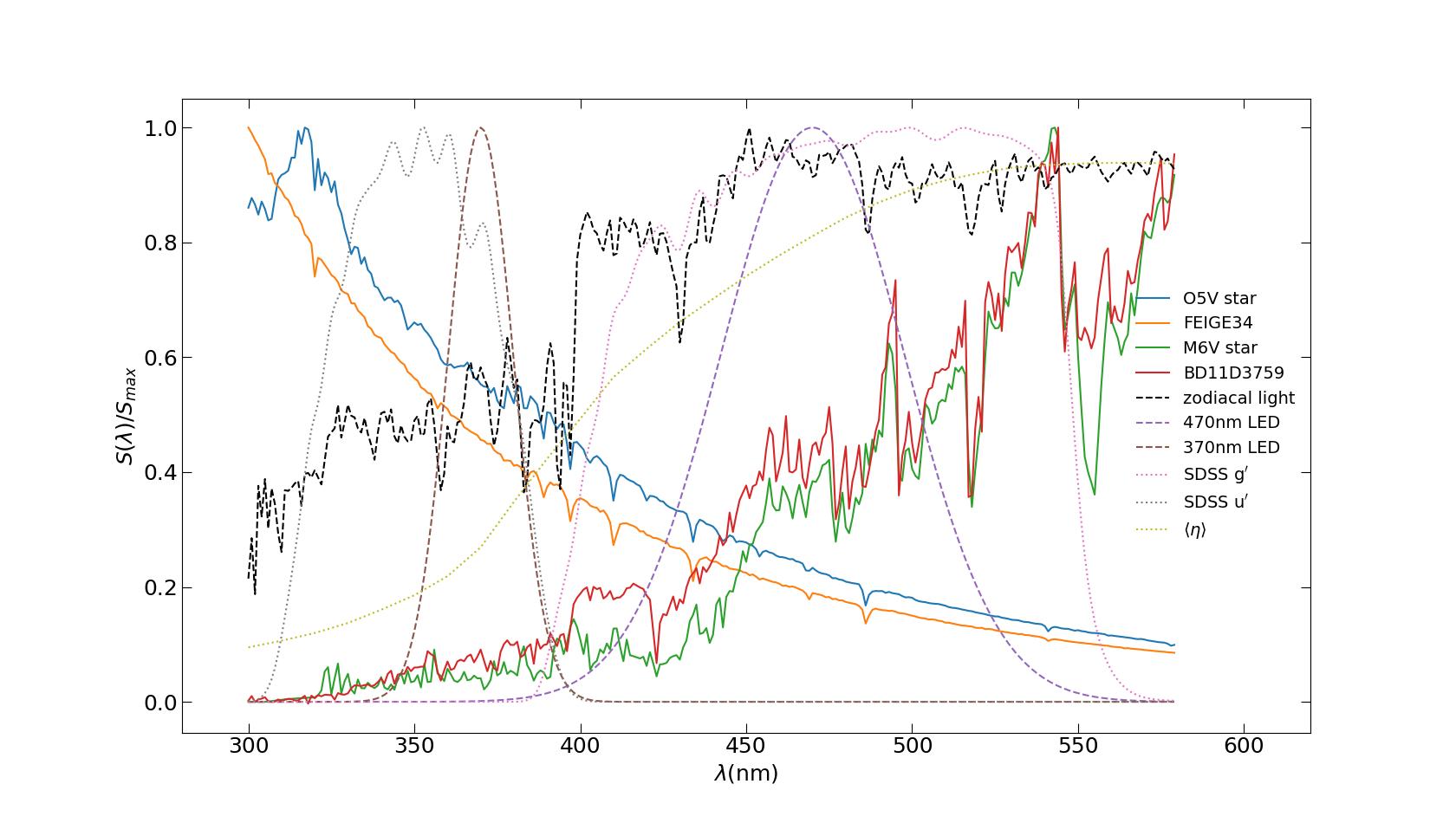}
\caption{SEDs of the target stars (O5V and M6V template), the standard stars (FEIGE and BD11D3759), the LEDs (370nm and 470nm) and the zodiacal light. Each curve is scaled by its maximum value in the plot range. SDSS u', g' filter transmission curves and the pixel-averaged quantum efficiency $\langle \eta \rangle$ from Figure~\ref{fig:fig3} are also shown in the plot.
\label{fig:fig9}}
\end{figure}

\begin{deluxetable}{cccccccc}[ht!]
\tablewidth{0pt} 
\tablehead{\colhead{\multirow{2}*{Spectral Type}} & \colhead{\multirow{2}*{band}} & \colhead{$\Delta m_\mathrm{sys}$} & \colhead{$\mathrm{RMS}_{n^\mathrm{t}-n^\mathrm{s}}$} & \colhead{\multirow{2}*{Flat Source}} & \colhead{$\mathrm{RMS}_{n^\mathrm{t}-n^\mathrm{f}}$} & \colhead{$\sigma_{\mathrm{rep}}$} & \colhead{$\sigma_{\mathrm{cal}}$}\\ & & \colhead{(mmag)} & \colhead{($10^{-3}$)} & & \colhead{($10^{-3})$} & \colhead{(mmag)} & \colhead{(mmag)}}
\tablecaption{Photometry errors of the target star in u' and g' bands with different flat-field illumination sources. \label{tab:magnitude_error}} 
\startdata
\multirow{6}*{O5V} & \multirow{3}*{u'} & \multirow{3}*{-12} & \multirow{3}*{0.20} & zodi & 0.86 & 1.1 & 1.3\\
  & & & & 354/23 & 2.0 & 2.7 & 2.6\\
 & & & & 370/23 & 2.0 & 2.7 & 2.7\\ \cline{2-8}
 & \multirow{3}*{g'} & \multirow{3}*{4} & \multirow{3}*{0.073} & zodi & 1.4 & 1.9 & 1.9\\
 & & & & 470/65 & 1.1 & 1.6 & 1.5\\
 & & & & 500/65 & 2.7 & 3.7 & 3.7\\
\hline
\multirow{6}*{M6V} & \multirow{3}*{u'} & \multirow{3}*{68} &  \multirow{3}*{0.58} & zodi & 1.1 & 1.5 & 1.9\\
 & & & & 354/23 & 3.9 & 5.2 & 5.6\\
 & & & & 370/23 & 1.5 & 2.1 & 2.2\\ \cline{2-8}
& \multirow{3}*{g'} & \multirow{3}*{-17} & \multirow{3}*{0.31} & zodi & 1.3 & 1.8 & 1.6\\
 & & & & 470/65 & 1.6 & 2.2 & 2.0\\
 & & & & 500/65 & 0.029 & 0.039 & 0.28
 \enddata
\tablecomments{In the ``Flat Source'' column, the zodiacal light is abbreviated as ``zodi'', and the LEDs are labelled in the form of ``$\lambda_\mathrm{c}$(nm)/FWHM(nm)''.}
\end{deluxetable}

As an example, we calculate the systematic error $\Delta m_\mathrm{sys}$ and the RMS photometric calibration error $\sigma_\mathrm{cal}$ with O5V and M6V template SEDs from UVKLIB library \citep{Pickles_1998} in SDSS u' band and g' band where the PRNU is either relatively large or changing rapidly with $\lambda$. The results are shown in Table \ref{tab:magnitude_error} 
along with the repeatability error $\sigma_\mathrm{rep}$.
Corresponding standard stars are chosen to be FEIGE34 and BD11D3759 from CALSPEC database \citep{2014PASP..126..711B}. We use the zodiacal light\footnote{Data from https://etc.stsci.edu/} and LEDs as flat-field illumination sources in the test and, hereafter, refer to the flats as zodi-flat and LED flat respectively. 
The LEDs' SEDs are approximated as Gaussians centered at various wavelengths $\lambda_\mathrm{c}$ (see Table \ref{tab:magnitude_error}).
Two LEDs with the same FWHM but slightly shifted $\lambda_\mathrm{c}$ are set for each filter band to show the sensitivity of the photometry errors to $\lambda_\mathrm{c}$. 
The LEDs' FWHMs here are similar to those in the dome flat system described in  \citet{Marshall_2013}.
The SEDs of the stars and the LEDs are illustrated in Figure~\ref{fig:fig9}. Also shown are the transmission curves of the two bands and the pixel-averaged QE $\langle \eta_i \rangle$ (same as the CCD QE curve in Figure \ref{fig:fig3}). The NFFs of various SEDs, $n^\mathrm{t}$, $n^\mathrm{s}$ and $n^\mathrm{f}$, are constructed using the best-fit parameters from Section \ref{subsec:section4.2}. The PSF is assumed to be a circular Gaussian with a FWHM of 5 pixels, large enough for most observations.
A smaller PSF would result in larger errors. 

The mismatch of the NFFs can be quantified by the RMS difference between the respective pairs of NFFs, \eg, $\mathrm{RMS}_{n^{\mathrm{t}}-n^{\mathrm{s}}}$ for $(n^\mathrm{t}-n^\mathrm{s})$ and $\mathrm{RMS}_{n^\mathrm{t}-n^\mathrm{f}}$ for $(n^\mathrm{t}-n^\mathrm{f})$.
By selecting the standard stars to be of the same type as the target stars, we expect their SED mismatch ($S^\mathrm{t}$--$S^\mathrm{s}$ mismatch) to be subdominant to the SED mismatch between the target/standard star and the flat field ($S^\mathrm{t}$--$S^\mathrm{f}$ mismatch) in terms of contribution to the calibration errors $\sigma_\mathrm{cal}$. 
Indeed, as \Tab{tab:magnitude_error} shows, $\mathrm{RMS}_{n^{\mathrm{t}}-n^{\mathrm{s}}}$ is well below $10^{-3}$ and is much smaller than $\mathrm{RMS}_{n^\mathrm{t}-n^\mathrm{f}}$ in all but one case. 
Moreover, the calibration error $\sigma_\mathrm{cal}$ is roughly equal to the repeatability error $\sigma_\mathrm{rep}$, which corroborates the notion that the $S^\mathrm{t}$--$S^\mathrm{f}$ mismatch is the main source of the calibration error in our tests. In real observations, however, the $S^\mathrm{t}$--$S^\mathrm{s}$ mismatch is not necessarily negligible.

Since the target objects can have wildly different SEDs, it is impossible to use one flat-field illumination source to match all. The RMS calibration error due to  $S^\mathrm{t}$--$S^\mathrm{f}$ mismatch is at a few mmag level and becomes an irreducible photometry error if one does not account for the wavelength-dependent PRNU. The broadband zodi-flat performs better than LED flats in most cases. 
Exceptions occur when the LED flat happens to produce a better matching $n^\mathrm{f}$ to $n^\mathrm{t}$, \ie{}, a smaller RMS$_{n^\mathrm{t}-n^\mathrm{f}}$, than the zodi-flat does.
For instance, in the extreme case of the M6V star in g' band, the 500nm LED gives more weight to long wavelengths, so qualitatively we expect it to work better than the roughly equal-weighting zodiacal light for the fairly ``red'' target SED. 
However, it is only a coincidence that the $S^\mathrm{t}$--$S^\mathrm{f}$ mismatch is negligible in this specific case.

\Tab{tab:magnitude_error} demonstrates that a broadband source such as the zodiacal light is a generally fail-safe choice for flat fielding if the error budget is at 0.01~mag level (excluding $\Delta m_\mathrm{sys}$). 
The margin becomes smaller with LEDs as the illumination source, which usually emit in rather narrow bands. 
Note that the errors in \Tab{tab:magnitude_error} correspond to the case with no attempt to correct for the wavelength-dependent PRNU. 
To suppress these errors, one may apply the PRNU model to construct the NFFs $n^\mathrm{t}$ and $n^\mathrm{s}$ and use them in place of $n^\mathrm{f}$ to do flat fielding for the target star and the standard star separately. 
The last term in \Eqn{eq28} then diminishes. 
In this case, one needs not only the standard star's SED but also the target's SED and accurate PRNU model parameters, which are not trivial to obtain. 
Nevertheless, the wavelength-dependent PRNU must be corrected for each object individually according to its SED, if mmag-level photometry precision is required. 

\section{Discussion and Conclusion} \label{sec:summary}
Image sensors' wavelength-dependent PRNU can introduce non-negligible photometry uncertainties that may not be fully corrected by routine flat fielding. In this paper, we use the four-parameter semi-physical 
PRNU model introduced by \citet{chen2018} to fit nine LED flat fields taken by a laser-annealed BSI CCD. Effects of photon absorption and photoelectron trapping or recombination on the PRNU are being modeled with an exponential decay probability. We propose a robust two-step fitting procedure, which also reduces the scale of the problem by 25$\%$. The results show that the flat fields can be reconstructed accurately with typical RMS errors less than 0.2\%. The RMS errors decrease further with the LSNU removed before model fitting. 

Given that the action of back-surface passivation for BSI CCDs is necessarily confined to the very top layer of the CCDs' back surface, we expect our method to be generically applicable to BSI CCDs, regardless of the back-surface passivation technique, though each technique can have its own distinct PRNU imprint or even featureless in the case of atomic layer deposition as we see on another CCD under development. 
The LSST camera actually has two types of CCDs installed: laser-annealed e2v CCD250 and chemisorption-coated ITL STA3800 \citep{Chemisorption_coating1998,STA3800}. CCD250 displays brick-wall patterns similar to those in \Fig{fig:fig1}, whereas STA3800 shows more complex patterns
\citep{LSSTCam_test2018,LSSTCam_PSF2023}. 
Our method can be particularly helpful to the forthcoming China Space Station Telescope \citep[CSST, also known as the Xuntian Space Telescope;][]{2011SSPMA..41.1441Z,Zhan2021,Gong_2019}, which plans to cover the wavelength range of 255--1000~nm with its multiband imaging and slitless spectroscopy survey. The CSST's high-resolution ($\le 0\arcsec.15$) NUV observations will be unique among surveys of its time and require a thorough characterization of its CCDs including the wavelength-dependent PRNU before launch.

\begin{figure}[ht]
\centering
\includegraphics[width=0.6\linewidth]{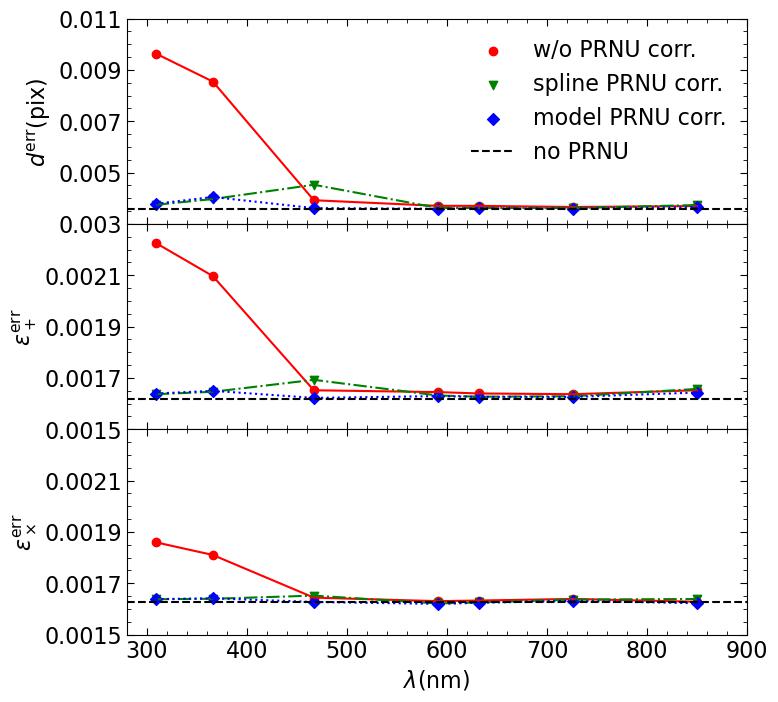}
\caption{
RMS errors of centroid position $d^\mathrm{err}$ and ellipticity components $\varepsilon_+^\mathrm{err}$ and $\varepsilon_\times^\mathrm{err}$ 
for a randomly placed circular Gaussian PSF with FWHM=3pix and SNR=300. The errors without PRNU correction (circles), those with correction by spline-interpolated NFFs (triangles) and those with correction by our model NFFs (diamonds) are shown along with those with no PRNU effect (dashed lines) for comparison.
\label{fig:astro_error}}
\end{figure}

It should be mentioned that not all mechanisms causing the PRNU features are included in our model. For example, we have not taken into account the redistribution of photoelectrons due to lateral electric fields in the CCD, which results in the tree-ring patterns and the brighter-fatter effect. We also neglect multiple reflections of long-wavelength photons within the CCD, which can increase the QE to some extent and may cause fringing for sufficiently narrow wavebands. It is expected that flat-field images can be reconstructed better by accounting for these additional effects. 

A potential application of our method is to correct the PRNU effect individually for each object according to its SED. As shown in Section~\ref{sec:error}, the SED difference between the target/standard star and the flat-field source causes mismatch in their PRNU patterns and introduces a small error in photometry according to \Eqn{eq28}. 
This effect can easily blow the error budget if one aims to achieve mmag level precision. With the wavelength-dependent PRNU model, one can predict the PRNU map or the NFF for any type of SED and reduce the error to an acceptable level. 
Section~\ref{subsec:section4.4} gives an example of such prediction. After removal of the flat field of one wavelength from the whole set of nine, the model fit to the remaining flat fields of eight wavelengths can generate the ``missing'' flat field with an RMS residual much less than 1\% in most cases. 
The accuracy can be further improved by taking flat fields at more wavelengths especially in the NUV, and  high-precision object-by-object PRNU correction can be achieved by combining SED fitting with our PRNU modeling approach. 

PRNU correction also affects astrometry and shape measurements. While it is quite involved to determine the position and ellipticity of an object or a PSF precisely \citep[e.g.,][]{2021MNRAS.501.1282J}, we provide a simple test here analogous to that in  Section~\ref{subsec:section4.4}.
We place a circular Gaussian PSF randomly in the LED flats from Section~\ref{sec:section2} and use the predicted flat fields in Section~\ref{subsec:section4.4} to correct for the PRNU. 
The PSF has an FWHM of 3 pixels and an SNR of 300. The SNR is set high enough to reveal the PRNU caused errors at 309nm and 366nm. 
Since the bandwidths of the LEDs are fairly narrow, we neglect the SED dependent effect in this test.
The PSF's centroid position and ellipticity components \citep[defined by quadruple moments, see \eg,][]{10.1093/mnras/sts371} are extracted using the {\sc sep} package \citep{SEP,1996A&AS..117..393B}.
\Fig{fig:astro_error} compares the centroid and ellipticity RMS errors of the PSF without PRNU correction (circles), those with correction by spline-interpolated NFFs (triangles), those with correction by our model NFFs (diamonds) and those with no PRNU effect (horizontal dashed lines). 
The ideal case of no PRNU effect sets the least measurement errors resulting from only the photon noise, readout noise and dark current, which we refer to as the statistical errors to distinguish them from those arising from the PRNU or PRNU correction.
It is assuring to see for such a high SNR case that even without correction the PRNU is still subdominant to the statistical errors at $\lambda \gtrsim 450$~nm and that our model fitting approach can suppress the errors at shorter wavelengths to the statistical level. 
The spline-interpolation method works well most of the time, though it sometimes introduces more errors instead of reducing them. 

The centroid errors of several thousandths of a pixel in \Fig{fig:astro_error} are small enough for most applications. For example, given the CSST pixel scale of $0\arcsec.074$, the position errors could be well below one microarcsecond in the absence of other errors.
Such precision is already comparable to that of the primary astrometric data set of Gaia Data Release 1 \citep{Gaia_DR1}, in which the sources were observed repeatedly for a medium of 15 times. 
In practice, errors from other sources are likely to swamp the errors caused by the PRNU correction and even exceed the statistical errors of bright stars at an SNR of a few hundred. 
The impact of the PRNU and its correction on ellipticity measurements is similar to that on position measurements in terms of the trend over wavelength. 
However, weak lensing studies place a very stringent requirement on the ellipticity model accuracy, \eg, RMS errors less than $5\times 10^{-5}$ for \emph{Euclid} \citep{refId0}. 
Even without any systematic effect, the statistical errors for SNR=300 stars as shown in \Fig{fig:astro_error} are already far from meeting the \emph{Euclid} requirement, and interpolation over the focal plane will introduce more errors.
The situation worsens with real observations, which rely on much fainter stars for PSF modeling \citep[see, \eg, simulation results in][]{refId0}, but we can still conclude that the PRNU and its correction are not a limiting factor for ellipticity measurements in this test.

The test above does not include the SED dependent effect discussed in Section~\ref{sec:error}, which might become significant especially for broadband observations at $\lambda\lesssim 450~\mathrm{nm}$.
In such a case, we expect that individual PRNU correction based on the object's SED can still suppress the errors caused by the wavelength-dependent PRNU to below the statistical level. 
It is however not trivial to implement, as, for most objects imaged in a large survey, one would not know their SEDs in advance. 
An iterative process of estimating the SED and correcting the PRNU appears necessary, and we plan to explore in this direction in the future.

\subsection*{ACKNOWLEDGEMENTS}
We acknowledge support by the National Key R\&D Program of China No. 2022YFF0503400 and the National Natural Science Foundation of China (NSFC) grant U1931208. W.D. is also supported by NSFC grants 11803043 and 11890691. 

\software{astropy \citep{2013AA...558A..33A,2018AJ....156..123A},  
          scipy \citep{scipy}, SEP \citep{SEP,1996A&AS..117..393B}}
          
\appendix
\section{Derivatives calculation}\label{sec:appendixA}
Partial derivatives of $S_1$ with respect to parameters of the $l$-th pixel $p_l = [P_l,d_l,\Delta_{l}]$ and that of $S_2$ with respect to $\delta_l$ are shown as follows
\begin{equation}
\begin{split}
\frac{\partial S_1}{\partial p_l} = \sum_k\sum_{j>k}\frac{\partial s_1(\lambda_j,\lambda_k)}{\partial p_l}\\
\frac{\partial s_1}{\partial p_l}(\lambda_j,\lambda_k) \equiv \frac{2}{\langle c \rangle}d_l(r_{1l}-\frac{A}{N\langle c \rangle})\\
c_l(\lambda_j,\lambda_k)\equiv t_l(\lambda_j)/t_l(\lambda_k)\\
d_l(\lambda_j,\lambda_k) \equiv \frac{1}{t_l(\lambda_{k})}[\frac{\partial t_l}{\partial p_l}(\lambda_j)-c_l(\lambda_j,\lambda_k)\frac{\partial t_l}{\partial p_l}(\lambda_k)]\\
r_{1l}(\lambda_j,\lambda_k) \equiv R_l(\lambda_j,\lambda_k)-R_l^{\mathrm{im}}(\lambda_j,\lambda_k)\\
A(\lambda_j,\lambda_k) \equiv \sum_l{r_{1l}(\lambda_j,\lambda_k)c_l(\lambda_j,\lambda_k)}\label{eqA1}
\end{split}
\end{equation}
\begin{equation}
\begin{split}
\frac{\partial S_2}{\partial \delta_l} = \sum_k\frac{\partial s_2(\lambda_k)}{\partial \delta_l}\\
\frac{\partial s_2}{\partial \delta_l}(\lambda) = \frac{2}{\langle \tau \rangle}[t_lr_{2l}+(t_1-t_l)\frac{B}{N\langle \tau \rangle}-r_{21}E_l]\\
r_{2l}(\lambda)\equiv f_l(\lambda)-f_l^{\mathrm{im}}(\lambda)\\
B(\lambda)\equiv \sum_{l\neq 1}r_{2l}(\lambda)\tau_l(\lambda)\\
E_l(\lambda)\equiv t_1(\lambda)+\frac{\tau_1(\lambda)}{N\langle \tau \rangle}[t_l(\lambda)-t_1(\lambda)]\label{eqA2}
\end{split}
\end{equation}\\
\begin{equation}
\begin{split}
t_l(\lambda) \equiv 1-e^{-H(1+\Delta_{l})/L(\lambda)}-\frac{P_ld_l}{L(\lambda)+d_l}\\
\frac{\partial t_l}{\partial P_l} = -\frac{d_l}{L+d_l}\\
\frac{\partial t_l}{\partial d_l} = -\frac{P_lL}{(L+d_l)^2}\\
\frac{\partial t_l}{\partial\Delta_{l}} = \frac{H}{L}e^{-H(1+\Delta_{l})/L}\label{eqA3}
\end{split}
\end{equation}
In \Eqn{eqA2} we have set $\delta_1 = -\sum_{l\neq 1}\delta_l$. $N=1024\times 1024$ is the number of pixels.

\section{Algorithm validity test}\label{sec:appendixB}
\begin{figure}[ht!]
\centering
\includegraphics[width=\linewidth]{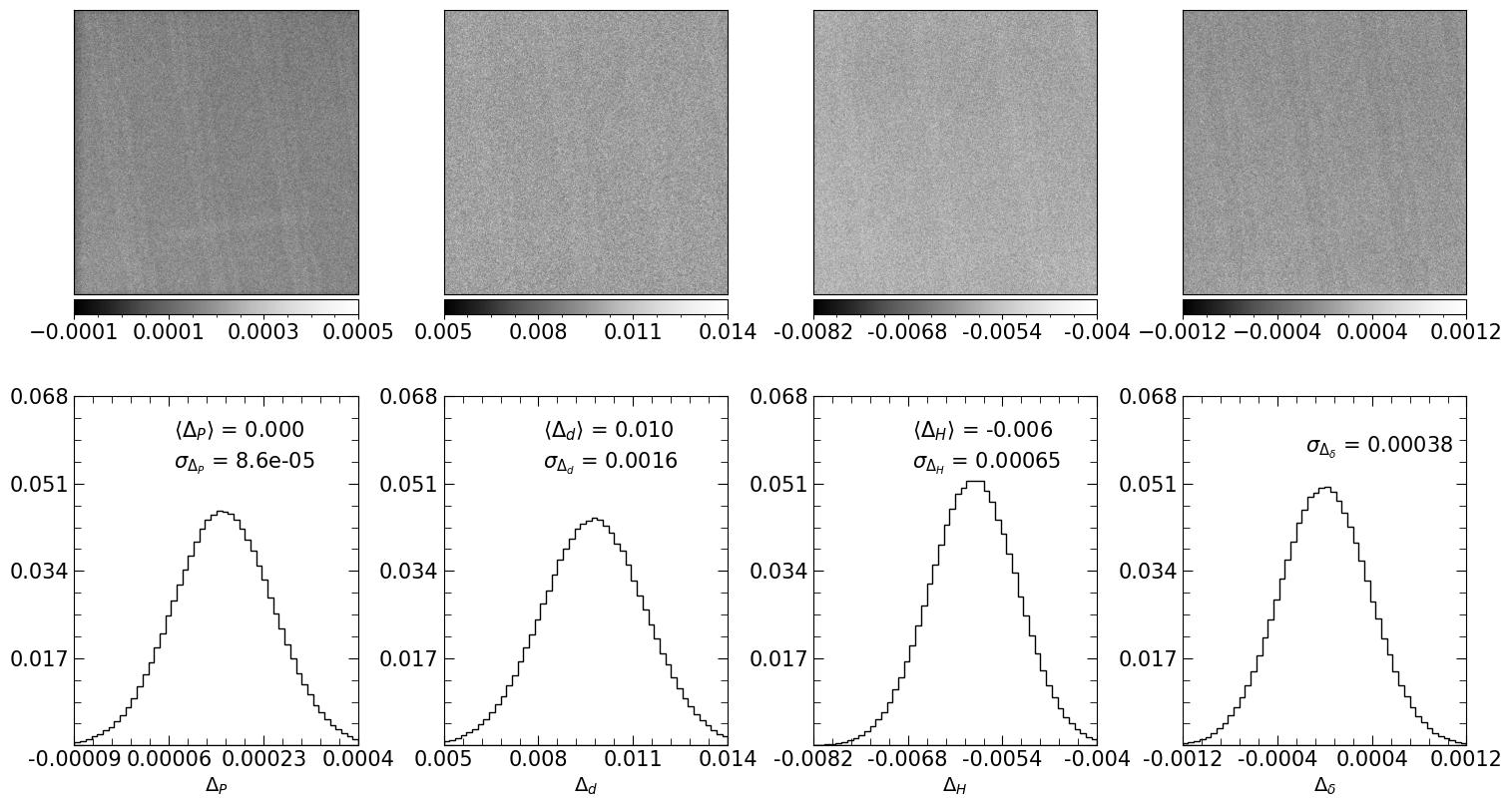}
\caption{Maps (top panel) and normalized histograms (lower panel) of synthetic image parameter fitting error distribution. $\Delta_P$, $\Delta_d$, $\Delta_H$, $\Delta_\delta$ are the relative error of $P$, $d$ and absolute error of $\Delta$, $\delta$ respectively. In the 1D histogram we show the average and standard deviation of parameter errors.    
\label{fig:fig14}}
\end{figure}

To validate the algorithm introduced in Section \ref{subsec:section4.1}, we generate mock flat images at nine wavelengths with photon noise according to our model and parameters shown in Figure \ref{fig:fig4}. We then employ the two-step optimization method and compare fitting results with input values. Maps and normalized histograms for error distribution of four parameters are shown in Figure \ref{fig:fig14}, where The definitions of fractional or absolute errors are given by $\Delta_{P}\equiv (P-P_{\mathrm{real}})/P_{\mathrm{real}}$, $\Delta_{d}\equiv (d-d_{\mathrm{real}})/d_{\mathrm{real}}$, $\Delta_{H}\equiv \Delta-\Delta_{\mathrm{real}}$ and $\Delta_{\delta}\equiv \delta-\delta_{\mathrm{real}}$ (subscript $\mathrm{real}$ represents real input parameters used for simulation). We can see that the shifts and scatters of reproduced $P$, $d$, $H$ compared with real values are all within one percent (absolute error of $\Delta_H$ can be considered as fractional error of $H$). For $\delta$ the scatter is shown to be a magnitude lower than the pixel scatter in Figure \ref{fig:fig4}. Thus our algorithm can effectively determine parameters of each pixel as long as our PRNU model is suitable for data images.

Fitting with millions of parameters might result in correlation between initial values and fitting results. To check if this problem exists in our algorithm, we fit the nine flat fields starting from five sets of uniformly drawn initial points within boundaries $0.4<P<1$, $0.05\mu\mathrm{m}<d<0.2\mu\mathrm{m}$, $-0.3<\Delta<0.3$ and $-0.1<\delta<0.1$. We calculate the largest parameter shift for each pixel, and the average over the CCD are 0.0023, 0.0016$\mu\mathrm{m}$, 0.00077 and $3.5\times 10^{-5}$ respectively for $P$, $d$, $\Delta$ and $\delta$. Shifts of $P$, $d$, $\Delta$ are negligible compared with their average over all pixels, and the shift of $\delta$ is significantly lower than its scatter. Thus we conclude that our model fitting algorithm is robust under different initial values.

\bibliography{paper}
\bibliographystyle{aasjournal}
\end{document}